%% file: mainDoc4.tex
\documentclass[journal,11pt,onecolumn,draftclsnofoot]{IEEEtran}

\usepackage{cite}

\ifCLASSINFOpdf
  \usepackage[pdftex]{graphicx}
\else
  \usepackage[dvips]{graphicx}
  \graphicspath{{../eps/}}
  \DeclareGraphicsExtensions{.eps}
\fi

\usepackage{tikz}
\usepackage{tikzscale}
\usepackage{pgf}
\usepackage[T1]{fontenc} 
\usepackage[pangram]{blindtext}

\usepackage{pgfplots}
\pgfplotsset{compat=newest}
\pgfplotsset{plot coordinates/math parser=false}
\newlength\figureheight
\newlength\figurewidth 

\usepackage{amsmath,amssymb}
\usepackage[numbers,sort&compress]{natbib}
\usepackage{algorithmic}

\usepackage{array}
\ifCLASSOPTIONcompsoc
  \usepackage[caption=false,font=normalsize,labelfont=sf,textfont=sf]{subfig}
\else
  \usepackage[caption=false,font=footnotesize]{subfig}
\fi

\usepackage{stfloats}
\usepackage{url}

\newcommand{\subparagraph}{}
\usepackage{titlesec}
\titlespacing*{\section}{1pt}{0.5\baselineskip}{0.5\baselineskip}
\titlespacing*{\subsection}{1pt}{0.45\baselineskip}{0.45\baselineskip}

\newtheorem{remark}{Remark}

\begin{document}
\title{Few-Bit CSI Acquisition for Centralized Cell-Free Massive MIMO with\\ Spatial Correlation}



\author{Dick Maryopi and Alister Burr
\\Department of Electronic Engineering\\
University of York, Heslington, York, UK.\\
Email: dm1110@york.ac.uk.

}

\maketitle
\begin{abstract}
The availability and accuracy of Channel State Information (CSI) play a crucial role for coherent detection in almost every communication system. Particularly in the recently proposed cell-free massive MIMO system, in which a large number of distributed Access Points (APs) is connected to a Central processing Unit (CPU) for joint decoding, acquiring CSI at the CPU may improve performance through the use of detection algorithms such as minimum mean square error (MMSE) or zero forcing (ZF). There are also significant challenges, especially the increase in fronthaul load arising from the transfer of high precision CSI, with the resulting complexity and scalability issues. In this paper, we address these CSI acquisition problems by utilizing vector quantization with precision of only a few bits and we show that the accuracy of the channel estimate at the CPU can be increased by exploiting the spatial correlation subject to this limited fronthaul load. Further, we derive an estimator for the simple \emph{Quantize-and-Estimate} (QE) strategy based on the Bussgang theorem and compare its performance to \emph{Estimate-and-Quantize} (EQ) in terms of Mean Squared Error (MSE).
Our simulation results indicate that the QE with few-bit vector quantization can outperform EQ and individual scalar quantization at moderate SNR for small numbers of bits per dimension.
\end{abstract}

\begin{IEEEkeywords}
Massive MIMO, Cell-Free, Fronthaul, Vector Quantization, Bussgang, Channel Estimation.
\end{IEEEkeywords}

\IEEEpeerreviewmaketitle

\section{Introduction}
\begin{figure}[!t]
\centering
\input{modelIlustration}
\caption{Illustration of Cell-Free Massive MIMO with $L$ access points, $N$ antenna per access point, $K$ users and finite fronthaul rate $C_l$.}
\label{systemmodel}
\end{figure}
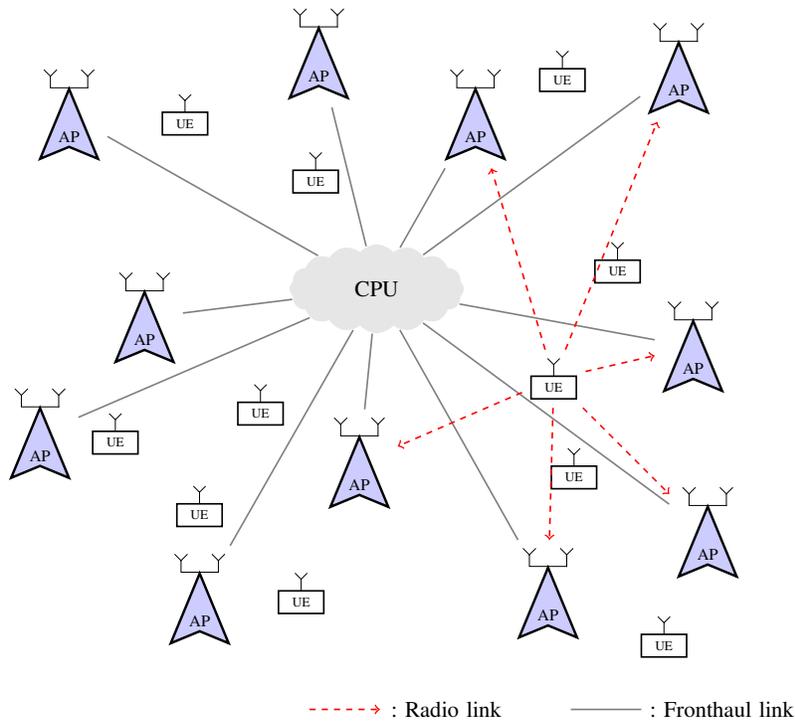

Cell-free massive MIMO is a new network architecture which has been gaining more attention recently as it has the potential to provide a high capacity per user and per unit area \cite{7827017, 8422577}. As illustrated in Fig. \ref{systemmodel}, the service area is not divided into separate cells, and the users are not associated with single base stations, but may communicate via multiple access points (APs) simultaneously. To enable this, the Access Points (APs) are distributed over the whole service area and are connected via fronthaul links to the Central Processing Unit (CPU) so that they can cooperate to detect user signals. 

Due to the large number of APs deployed in Cell-Free massive MIMO, distributed detection and precoding are often performed at the APs thus reducing the complexity and improving the scalability of channel state information (CSI) acquisition at the CPU. It also avoids the additional fronthaul load due to CSI transfer. However, this approach sacrifices significant achievable data rates in contrast to the centralized approach where the CSI is available at the CPU \cite{8422865, 8417560, TVTcorrMBB}. In this case (on the uplink) maximum ratio combining (MRC) must be performed at the APs, and separately weighted estimates of each user's signal transmitted to the CPU.  It is shown in [2] that the overall fronthaul load may then be higher, and then overall performance poorer, than the case where the quantized CSI and quantized user signals are transferred to the CPU. Moreover if the CSI is available at the CPU more effective detection algorithms such as zero forcing (ZF) may be used, further improving performance [3].

A natural question is therefore whether one can provide CSI at CPU with low fronthaul requirement and with relative simple CSI acquisition. While there has been a number of studies about CSI acquisition and fronthaul load reduction in the context of Cloud Radio Access Networks (C-RAN) \cite{7444125}, there are few on Cell-Free massive MIMO \cite{TVTcorrMBB, 8445960}. For instance, the CSI acquisition strategies taking into account the limited fronthaul capacity for single-antenna APs has been investigated in \cite{TVTcorrMBB}. In the case of multiple-antenna APs the work in \cite{8445960} studied CSI acquisition at the CPU by utilizing sophisticated hybrid beamforming, where analog combining is performed prior to one-bit analog-to-digital conversions (ADCs).

Nevertheless, the above works do not explicitly address the channel correlation. For one-bit co-located massive MIMO the problem of channel estimation with correlation was studied among others in \cite{Kim2018ChannelEF}. Since the real channel tends to be spatially correlated, in this paper we study CSI acquisition for cell-free massive MIMO with limited fronthaul capacity and assuming spatial correlation at multi-antenna APs. In this case, we use Vector Quantization (VQ) with precision of only a small number of bits, so as simultaneously to exploit the channel correlation and meet the low bit requirement of the fronthaul. By applying a vector quantization at APs we investigate the CSI acquisition strategy \emph{Quantize-and-Estimate} (QE) based on the Bussgang theorem and its counterpart \emph{Estimate-and-Quantize} (EQ). Our simulation results demonstrate that this few-bit vector quantization can exploit the channel correlation effectively giving better performance in terms of Mean Squared Error (MSE) compared to utilizing Scalar Quantization (SQ) at individual antennas. Further, the QE scheme, which is relatively simple for the implementation at the APs, is shown to offer a significant performance improvement over EQ particularly at low to moderate SNR.

The rest of the paper is organized as follows. In section \ref{SM} the system model is described including the spatial channel correlation model. We describe then our considered fronthaul compression technique in section \ref{FC}. In section \ref{CSI A}, we present our CSI acquisition schemes, where we make use of the vector quantization and Bussgang decomposition described in section \ref{FC}. We evaluate then our schemes numerically in section \ref{NR} and close with conclusion in section \ref{Conclusion}.

\emph{Notation}: Roman letter, lower-case boldface letter and upper-case boldface letter are used respectively to denote a scalar, a column vector and a matrix. The set of all complex and real $M\times N$ matrix are represented by $\mathbb{R}^{M \times N}$ and $\mathbb{C}^{M \times N}$ respectively. By $\langle \cdot, \cdot\rangle$ we denote the inner product with $\Vert \cdot \Vert$ as its corresponding vector norm or Frobenius norm. The expectation of random variable is represented by $\mathbb{E}\{\cdot\}$. We denote circularly complex Gaussian distribution with mean $\mathbf{m}$ and covariance matrix $\mathbf{\Sigma}$ by $\mathcal{CN}(\mathbf{m}, \mathbf{\Sigma})$. We use $\mathbf{I}_N$ for $N\times N$ identity matrix and $\mathbf{1}_N$ for all-one vector of dimension $N$. We denote the transpose conjugate by $(\cdot)^H$. For a vector $\mathbf{a}$, $\text{diag}(\mathbf{a})$ denotes a diagonal matrix with the diagonal elements created from vector $\mathbf{a}$.

\section{System Model} \label{SM}
We consider uplink transmission in a cell-free system \cite{7827017}, where we have $K$ single-antenna users (UEs) and $L$ Access Points (APs) equipped with $N\geq 1$ antennas. We fix the total number of AP antennas in the system at $M=LN$. The processing of the signals received at the APs is virtualized at the Central Processing Unit (CPU), which is connected to the L APs by $L$ error-free fronthaul links which carry the signals in digitally encoded form.  
  
\subsection{Channel Model}
We denote the channel between the $k$-th user and the $m$-th antenna of the $l$-th AP by $g_{mk}$ where $m=(l-1)N+1, \dots, lN$ for $l=1,\dots, L$, and $k=1, \dots, K$. For a given $l$ and $k$ the channel is specified by the $N\times 1$ vector $\mathbf{g}_{lk}\sim\mathcal{CN}(\mathbf{0}_N, \mathbf{\Sigma}_{lk})$ where $\mathbf{\Sigma}_{lk}\in\mathbb{C}^{N \times N}$ is the covariance matrix including the large scale fading and the spatial correlation given by
\begin{align}
\mathbf{\Sigma}_{lk}=\beta_{lk}\mathbf{R}_{lk}.
\end{align} 
The large scale fading $\beta_{lk}$ is a path-loss dependent coefficient whereas the correlation matrix $\mathbf{R}_{lk}\in\mathbb{C}^{N \times N}$ is dependent on the particular environment between AP and UE. In this case, we follow the local scattering model given in \cite{SIG-093}, where any user $k$ at the azimuth angle $\theta$ to the AP $l$ is surrounded by scatterers causing correlation between the multipath signal components received between at the antennas of the AP. Accordingly, the correlation coefficient can be specified by an angle of arrival $\bar{\theta}$ which is treated as a random variable with probability density function $f(\bar{\theta})$ and the entries of the correlation matrix $\mathbf{R}_{lk}$ are then determined by
\begin{align}
\left[\mathbf{R}_{lk}\right]_{a,b}= \int e^{j2\pi d_H(a-b) sin(\bar{\theta})} f(\bar{\theta}) d\bar{\theta},
\end{align}
where $d_H$ is the spacing between antennas $1\leq a, b \leq N$. Further, $\bar{\theta}$ can be expressed as $\bar{\theta}=\theta+\delta$, where $\delta$ is a random angular spread with standard deviation $\sigma_\delta$.

Using the Karhunen-Loeve representation we can describe the correlated channel vector as
\begin{equation}
\mathbf{g}_{lk}=\beta_{lk}^{1/2}  \mathbf{U}_{l} \Lambda_l^{1/2}  \mathbf{h}_{lk} , \label{gmk}
\end{equation}
where the vector $\mathbf{h}_{lk}\sim\mathcal{CN}(\mathbf{0}_N, \mathbf{I_{N}})$ models the small scale fading between the $k$-th user and the $l$-th AP. The unitary matrix $\mathbf{U}\in\mathbb{C}^{N \times r}$ and the diagonal matrix $\Lambda\in\mathbb{R}^{r \times r}$ comprise respectively the eigenvectors and the associated eigenvalues of the correlation matrix $\mathbf{R}_{lk}$ with rank $r$. The channel vector of the $k$-th user to all $L$ APs is then given by $\mathbf{g}_{k}\sim\mathcal{CN}(\mathbf{0}_M, \mathbf{\Sigma}_{k})$, where $\mathbf{\Sigma}_{k}=\text{diag}\left(\mathbf{\Sigma}_{1k}, \dots,\mathbf{\Sigma}_{Lk}\right)$. Further, we stack the channel from $K$ users to all $L$ APs in the columns of the $M\times K$ matrix $\mathbf{G}=\left[\mathbf{g}_{1}, \dots, \mathbf{g}_{K}\right]$, such that under the assumption of perfect fronthaul the received signal at the CPU can be modeled as
\begin{align}
\mathbf{y}=\mathbf{G}\mathbf{x}+\mathbf{w},
\end{align} 
where $\mathbf{x}\in\mathbb{C}^{K}$ is the channel input from all $K$ users and $\mathbf{w}\sim\mathcal{CN}(\mathbf{0}_M, \mathbf{I}_{M})$ is the i.i.d. additive Gaussian white noise at APs. Later, we will remove the assumption of perfect fronthaul and assume that the $l$-th fronthaul link connecting the $l$-th AP to the CPU can transmit quantized signals reliably at a maximum rate of $C_l$.

\section{Fronthaul Compression} \label{FC}
Due to the limited capacity of the fronthaul link and the high load of the digitally encoded signal we need to compress this data for efficient transmission to the CPU. To simplify our analysis, we consider fronthaul links with equal capacity of $C_l=C$ bits for all $l \in \{1,\dots,L\}$. 

\subsection{Vector Quantization}
Our considered compression consists of vector quantization followed by fixed-rate lossless coding. At each AP a vector quantizer $Q$ is applied as interface to the fronthaul with 
\begin{align}
Q(\mathbf{x})= \sum_{i=0}^{\mathcal{S}-1} q_i T_i(\mathbf{x}), \text{ where } 
T_i(\mathbf{x})=
\begin{cases}
1& \text{ if } \mathbf{x}\in\mathcal{C}_i\\
0& \text{ otherwise. } 
\end{cases} \label{VQ}
\end{align}
Whenever the input vector $\mathbf{x}\in\mathbb{R}^N$ falls into the cell $\mathcal{C}_i$, the index i will be transmitted on the fronthaul link, and the reconstruction value $q_i$ taken from the codebook $\mathcal{Q}=\{q_i\}_{i=0}^{\mathcal{S}-1}\subset\mathbb{R}^N$ will be used at the CPU. The codebook size corresponds to the fronthaul capacity  by $\mathcal{S}=2^{C}$. For $N$-dimensional vector quantization we allocate $C/N$ bits per dimension. Here, we keep $C/N$ small, to one or two bits per dimension. For a complex-valued signal we quantize separately the real and imaginary part. We do this for the reason that the correlation affects the real and imaginary part of the channel independently.

The optimal codebooks can be found using the Linde Buzo Gray (LBG) algorithm for minimum mean squared error \cite{Gersho:1991:VQS:128857}. This algorithm is the counter part of Lloyd algorithm for vector quantization, where the optimal codebook is obtained by alternating between finding the optimal partition by the nearest neighbour criterion and finding the optimal reconstruction values by the centroid condition. The contribution of our scheme compared to separate quantization is that we take the received signal from $N$ antennas at the AP as the input $\mathbf{x}$ of our vector quantizer $Q$. By doing so we expect to adapt the codebooks to the spatial channel correlation.

\subsection{Bussgang Decomposition}
The quantizer $Q$ given in (\ref{VQ}) is in general non-linear and the error $\mathbf{e}\triangleq\mathbf{x}-Q(\mathbf{x})$ resulting from the quantization process is correlated with the input vector $\mathbf{x}$. However, using the Bussgang theorem \cite{Bussgang52} we can express our quantizer as the following linear model
\begin{align}
\mathbf{x}_q=Q(\mathbf{x})=\mathbf{F}\mathbf{x}+\mathbf{d}, \label{LinearBussgang}
\end{align}
where for a Gaussian input the distortion $\mathbf{d}$ is statistically equivalent to the quantization error $\mathbf{e}$ but uncorrelated with the signal component $\mathbf{x}$. The linear operator $\mathbf{F}$, which depends essentially on the given distortion characteristic of $Q$, tells us also about the proportional factor between the input-output covariance of the quantizer expressed as \cite{Bussgang52}
\begin{align}
\mathbf{C}_{xx_q}&=\mathbf{F}\mathbf{C}_{xx}, \text{ where } \\
\mathbf{C}_{xx_q}&=\mathbb{E}\{\mathbf{x}\mathbf{x}_q^H\} \text{ and } \mathbf{C}_{xx}=\mathbb{E}\{\mathbf{x}\mathbf{x}^H\}.
\end{align}
In this case, finding $\mathbf{F}$ can be seen as finding the LMMSE estimator for $\mathbf{x}_q$ from the observation $\mathbf{x}$ \cite{capacityQuantMIMO}
\begin{align}
\mathbf{F}&=\mathbf{C}_{xx_q}\mathbf{C}_{xx}^{-1}, \label{F}
\end{align}
where the estimation error $\mathbf{d}$ is then orthogonal to $\mathbf{x}$. Using (\ref{F}) the covariance of the distortion $\mathbf{d}$ can also be expressed as 
\begin{align}
\mathbf{C}_{dd}&=\mathbb{E}\{(\mathbf{x}_q-\mathbf{F}\mathbf{x})(\mathbf{x}_q-\mathbf{F}\mathbf{x})^H\}\nonumber\\
&=\mathbf{C}_{x_qx_q}-\mathbf{C}_{x_qx}\mathbf{C}_{xx}^{-1}\mathbf{C}_{xx_q}.
\end{align}
The closed form expression of $\mathbf{F}$ is not yet known for a general quantizer, particularly for vector quantizers. Therefore, we compute $\mathbf{F}$ numerically whenever it is needed by assuming that we have access to measurements of the input as well as the output of $Q$. We estimate the covariance matrix $\mathbf{C}_{xx}$ from the sample covariance matrix
\begin{align}
\hat{\mathbf{C}}_{xx}=\frac{1}{N_{t}}\sum_{n_{t}=1}^{N_{t}} \mathbf{x}[n_{t}]\mathbf{x}[n_{t}]^H \label{SampleCovarince}
\end{align}
and respectively for $\mathbf{C}_{x_qx_q}$ and $\mathbf{C}_{xx_q}$. The number of observations $N_{t}$ can be conveniently taken equally to the number of codebooks' training, where $\hat{\mathbf{C}}_{xx}$ will approach $\mathbf{C}_{xx}$ for large $N_{t}$.

\section{CSI Acquisition Strategies}	\label{CSI A}
\begin{figure*}[ht!]\centering
\includegraphics[width=\textwidth]{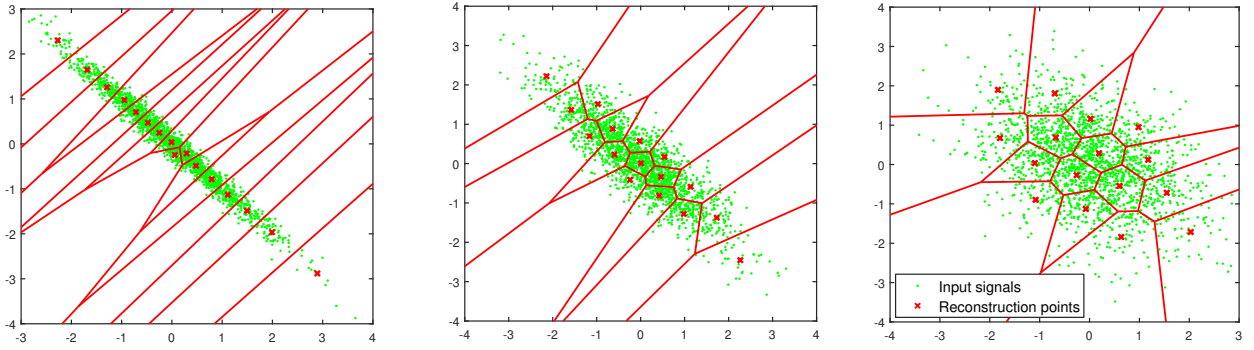} 
\caption{The voronoi region of codebook $\mathcal{Q}$ for $N=2$ and for different degree of correlation (i.e. for different angular spread standard deviation $\sigma_{\delta}$ of Gaussian distributed $\delta$).}
\label{voronoiFig}
\end{figure*}
We assume in this work that neither the APs nor the CPU knows a priori the channel realization $\mathbf{G}$. To enable interference suppression at the CPU, the CSI is needed at the CPU. We consider two strategies: estimate and quantize, and quantize and estimate, which will be discussed in the next two sub-sections.  

\subsection{Estimate-and-Quantize}
In this scheme we estimate first the channel at APs and then quantize the resulting CSI with the quantizer given in (\ref{VQ}) to meet the fronthaul-capacity limit of $C$ bits. The channel estimation is done based on the transmission of known pilots. Every user uses a specific sequence taken from a set $\Psi$ of orthonormal random sequences $\varphi_k \in\mathbb{C}^{\tau\times 1}$ with $\langle\varphi_k, \varphi_k'\rangle=\delta_{kk'}$ and $\Vert\varphi_k\Vert ^2=1$, where the sequence length $\tau$ is assumed to be less or equal than the coherence interval $\tau_c$. The $k$-th user sends $\sqrt{\tau}\varphi_k$ as its pilot such that the $l$-th AP observes the received pilot $\mathbf{Y}_{p,l}\in\mathbb{C}^{N \times \tau}$ from all $K$ users as
\begin{equation} \label{receivePilot}
\mathbf{Y}_{p,l}= \sqrt{\tau\rho_p}\sum_{k=1}^{K} \mathbf{g}_{lk}\varphi_k^H + \mathbf{W}_l,
\end{equation}
where $\rho_p$ is the transmit SNR of the pilot and $\mathbf{W}_l$ is an additive noise matrix at the $l$-th AP whose entries are uncorrelated with zero mean and unit variance. 

To allow all pilots to be orthogonal for all $K$ users, only $K\leq \tau$ users may transmit their pilots simultaneously. In this case, the transmitted pilots satisfy
\begin{equation}
\Phi^H\Phi=\tau\rho_p \mathbb{I}_K,\quad \text{ where } \Phi=\sqrt{\tau\rho_p}[\varphi_1, \dots,\varphi_K]. \label{pilotMatrix}
\end{equation}

The channel vector $\mathbf{g}_{lk}$ can be estimated at the APs where the received pilot $\mathbf{Y}_{p,l}$ is projected onto $\varphi_k$ expressed as
\begin{align}
\mathbf{r}_{p,lk}&=\frac{1}{\sqrt{\tau\rho_p}}\mathbf{Y}_{p,l}\varphi_k\nonumber\\
&=\mathbf{g}_{lk}+\sum_{k'\neq k}^{K} \mathbf{g}_{lk'}\varphi_{k'}^H\varphi_{k} + \frac{1}{\sqrt{\tau\rho_p}}\mathbf{W}_l\varphi_{k} \label{projectPilot}
\end{align}
To obtain the estimate of $\mathbf{g}_{lk}$ we use the LMMSE estimator given by 
\begin{align}
\hat{\mathbf{g}}_{lk}= \mathbf{\Gamma}_{lk}\mathbf{r}_{p,lk} \label{EstimatorIdeal} 
\end{align}
The gain matrix $\mathbf{\Gamma}_{lk}$ is given by 
\begin{align}
\mathbf{\Gamma}_{lk}&=\mathbf{\Sigma}_{lk}\left(\mathbf{\Omega}_{lk}\right)^{-1}, \text{ where }
\mathbf{\Sigma}_{lk}=\mathbb{E}\{\mathbf{g}_{lk}\mathbf{g}_{lk}^H\} \text{ and } \\
\mathbf{\Omega}_{lk}&=\mathbb{E}\{\mathbf{r}_{p,lk}\mathbf{r}_{p,lk}^H\}=\mathbf{\Sigma}_{lk}+\frac{1}{\tau\rho_p}\mathbf{I}_{N}
\end{align}

After accomplishing the channel estimation the APs quantize the channel estimate $\hat{\mathbf{g}}_{lk}$. We assume that the large scale fading $\beta_{lk}$ is relatively constant over a long period and known at the APs. Thus, we may scale the input to the vector quantizer accordingly with $\beta_{lk}$ and approximate the distribution as multivariate Gaussian. Consequently, we can optimize the codebook $\mathcal{Q}$ for each AP off-line and need only update it as the $\beta_{lk}$ changes. As demonstrated in Fig. \ref{voronoiFig} for $N=2$ the codebooks can exploit the spatial channel correlation effectively. Due to the LBG algorithm the reconstruction points $\{q_i\}$ are placed more densely in the region where the input signals come with high probability. As the correlation increases, the reconstruction points get closer to the diagonal to optimally represent the dependency between input signals. Thus, the distance from the input signals to the points $\{q_i\}$ becomes smaller resulting a smaller average distortion.

\subsection{Quantize-and-Estimate}
Instead of transferring the quantized CSI we consider here another CSI acquisition strategy where we quantize first the received pilots at the APs and then estimate the channel from the quantized pilots at the CPU. To be more specific, the $l$-th AP quantizes the receive pilots at the $N$ antennas jointly as
\begin{align}
\mathbf{y}_{qp,l}^{(t)}=Q(\mathbf{y}_{p,l}^{(t)}) 
&=Q\left(\sqrt{\tau\rho_p}\sum_{k=1}^{K} \mathbf{g}_{lk}{\varphi_k^{(t)}}^* + \mathbf{w}_l^{(t)} \right)\nonumber \\
&=Q\left(\sqrt{\tau\rho_p}\mathbf{G}_l{\varphi^{(t)}}^H + \mathbf{w}_l^{(t)} \right) \label{quantPilots}
\end{align}
where the superscript $t=\{1, \dots, \tau\}$ denotes the index of the pilot sequence. Accordingly $\mathbf{y}_{p,l}^{(t)}$ is the $t$-th column of $\mathbf{Y}_{p,l}$ in (\ref{receivePilot}) and $\varphi^{(t)}$ is the $t$-th row of $\Phi$ in (\ref{pilotMatrix}). 

Applying the Bussgang decomposition to (\ref{quantPilots}) we obtain
\begin{align}
\mathbf{y}_{qp,l}^{(t)}&=\mathbf{F}_{p,l}\mathbf{y}_{p,l}^{(t)} + \mathbf{d}_{p,l}^{(t)}\nonumber \\
&=\sqrt{\tau\rho_p}\mathbf{F}_{p,l}\mathbf{G}_l{\varphi^{(t)}}^H+\mathbf{F}_{p,l}\mathbf{w}_l^{(t)} + \mathbf{d}_{p,l}^{(t)}. \label{y_qp.l^(t)}
\end{align}
The CPU receives from all $L$ APs as a stack of (\ref{y_qp.l^(t)})
\begin{align}
\mathbf{y}_{qp}^{(t)}=
\begin{bmatrix}
\mathbf{y}_{qp,1}^{(t)}\\
\vdots \\
\mathbf{y}_{qp,L}^{(t)}
\end{bmatrix}
=\begin{bmatrix}
\sqrt{\tau\rho_p}\mathbf{F}_{p,1}\mathbf{G}_1{\varphi^{(t)}}^H+\mathbf{F}_{p,1}\mathbf{w}_1^{(t)} + \mathbf{d}_{p,1}^{(t)}. \\
\vdots \\
\sqrt{\tau\rho_p}\mathbf{F}_{p,L}\mathbf{G}_L{\varphi^{(t)}}^H+\mathbf{F}_{p,L}\mathbf{w}_L^{(t)} + \mathbf{d}_{p,L}^{(t)}.
\end{bmatrix}
\end{align}
where we can concisely rewrite as a $M\times \tau$ matrix for $\tau$-length sequences of quantized received pilots given by
\begin{align}
\mathbf{Y}_{qp}=
\begin{bmatrix}
\mathbf{y}_{qp,1}^{(1)}& \dots  &\mathbf{y}_{qp,1}^{(\tau)}\\
\vdots & \ddots & \vdots\\
\mathbf{y}_{qp,L}^{(1)} & \dots & \mathbf{y}_{qp,L}^{(\tau)}
\end{bmatrix}
=\sqrt{\tau\rho_p}\mathbf{\tilde{F}}\mathbf{G}{\Phi}^H+\mathbf{\tilde{F}}\mathbf{W} + \mathbf{D},
\end{align}
where the matrix $\mathbf{\tilde{F}}$ is a $M\times M$ diagonal matrix with $\mathbf{F}_{p,l}\in \mathbb{R}^{N\times N}$ in its  block diagonal entries. With similar structure to $\mathbf{Y}_{qp} \in \mathbb{C}^{M\times \tau}$ the matrix $\mathbf{W}$ and $\mathbf{D}$ denote respectively the noise and distortion matrix.

We can then project $\mathbf{Y}_{qp}$ onto $\varphi_k$ expressed as
\begin{align}
\mathbf{r}_{qp,k}&=\frac{1}{\sqrt{\tau \rho_p}}\mathbf{Y}_{qp}\varphi_k \nonumber\\
&=\mathbf{\tilde{F}}\mathbf{g}_{k}+\mathbf{\tilde{F}}\sum_{k'\neq k}^{K} \mathbf{g}_{k'}\varphi_{k'}^H\varphi_{k} + \frac{1}{\sqrt{\tau\rho_p}}\left(\mathbf{\tilde{F}}\mathbf{W}+\mathbf{D}\right)\varphi_{k}. 
\end{align}
We estimate $\mathbf{g}_{k}$ using LMMSE estimator
\begin{align}
\hat{\mathbf{g}}_{qp, k}= \mathbf{\Gamma}_{qp,k}\mathbf{r}_{qp,k}. \label{EstimatorBussgang}
\end{align}
The gain matrix $\mathbf{\Gamma}_{qp,k}$ is given by 
\begin{align}
\mathbf{\Gamma}_{qp,k}&=\mathbf{\Sigma}_{k}\mathbf{\tilde{F}}^H\left(\mathbf{\Omega}_{qp,k}\right)^{-1}, \text{ where }
\mathbf{\Sigma}_{k}=\mathbb{E}\{\mathbf{g}_{k}\mathbf{g}_{k}^H\} \text{ and } \nonumber \\
\mathbf{\Omega}_{qp,k}&=\mathbb{E}\{\mathbf{r}_{qp,k}\mathbf{r}_{qp,k}^H\}=\mathbf{\tilde{F}}\mathbf{\Sigma}_{k}\mathbf{\tilde{F}}^H+\frac{1}{\tau\rho_p}\left(\mathbf{\tilde{F}}\mathbf{\tilde{F}}^H+\mathbf{D}\mathbf{D}^H\right)
\end{align}

\begin{remark}
We assume that the received signals at the APs are uncorrelated over $l$ and $t$ such that the Gram matrix $\mathbf{\tilde{F}}\mathbf{\tilde{F}}^H$ and $\mathbf{D}\mathbf{D}^H$ have a block diagonal structure. Further, their submatrices are positive definite since $\mathbf{F}$ and $\mathbf{d}$ in (\ref{LinearBussgang}) are positive definite for large observation in sample covariance matrix (\ref{SampleCovarince}). Thus, the matrix $\mathbf{\Omega}_{qp,k}$ is invertible.
\end{remark}
%
\section{Numerical Results} \label{NR}
We provide in this section some numerical simulations to asses the performance of the considered schemes above. We did our simulation with $M=120$ total number of the APs' antennas, $L=120/N$ APs and $K=20$ number of users distributed uniformly in the area of $1\times 1 \text{ km}^2$. This area is wrapped around by its copies so that it resembles a network with infinite area. The channel $\mathbf{g}_{lk}$ in (\ref{gmk}) is modeled with the large scale fading $\beta_{lk}$ given as
\begin{align}
\beta_{lk}=\text{PL}_{lk}\cdot 10^{\frac{\sigma_{sh}z_{lk}}{10}},
\end{align}
where the factor $10^{\frac{\sigma_{sh}z_{lk}}{10}}$ is the uncorrelated shadowing with the standard deviation $\sigma_{sh}= 8 \text{ dB}$ and $z_{lk}\sim \mathcal{N}(0, 1)$. The path loss coefficient follows the three-slope model according to 
\begin{align}
\text{PL}_{lk}=
\begin{cases}
-\mathcal{L}-35\text{log}_{10}(d_{lk}), d_{lk} > d_1 \\
-\mathcal{L}-15\text{log}_{10}(d_1)-20\text{log}_{10}(d_{lk}), d_0<d_{lk}\leq d_1  \\
-\mathcal{L}-15\text{log}_{10}(d_1)-20\text{log}_{10}(d_0), d_{lk}\leq d_0 
\end{cases}
\end{align}
where $d_{lk}$ is the distance between the $l$-th AP and the $k$-th user, $d_0=0.01\text{ km}, d_1=0.05\text{ km}$, and
\begin{align}
\mathcal{L}&\triangleq 46.3+33.9 \text{log}_{10}(f)-13.83 \text{log}_{10}(h_{AP})\nonumber\\
&-(1.1 \text{log}_{10}(f)-0.7)h_u+(1.56 \text{log}_{10}(f)-0.8).
\end{align}
We choose the carrier frequency $f=1.9\text{ GHz}$, the AP antenna height $h_{AP}=15\text{ m}$ and the user antenna height $h_u=1.65\text{ m}$. For all user pilots we set the normalized transmit SNR $\rho_p$ equal, that is, the transmit power divided by the $\text{noise power} = B\times k_b\times T_0 \times \text{noise figure}$, where the bandwidth $B=20\text{ MHz}$, the Boltzmann constant $k_b=1.381 \times 10^{-23}$, the noise temperature $T_0=290$ Kelvin and the noise figure $=9$ dB.

In this paper, we use the MSE as the performance metric, defined as
\begin{align}
\text{MSE}= \frac{1}{M K}\mathbb{E}\{\Vert \mathbf{G}-\hat{\mathbf{G}} \Vert ^2\}, 
\end{align}
where $\hat{\mathbf{G}}$ is the channel matrix estimate depending on the employed acquisition scheme. The MSE of different schemes is evaluated by Monte Carlo simulation, where the transmission of orthogonal pilots of length $\tau=K=20$ is repeated over a sufficient number of independent realizations. For each large scale fading realization we carry out off-line  training with $N_{t}=100$ over random small scale fading to approach the optimal codebooks for our vector quantizers. 
\begin{figure}[ht!]\centering
\includegraphics[width=0.7\columnwidth]{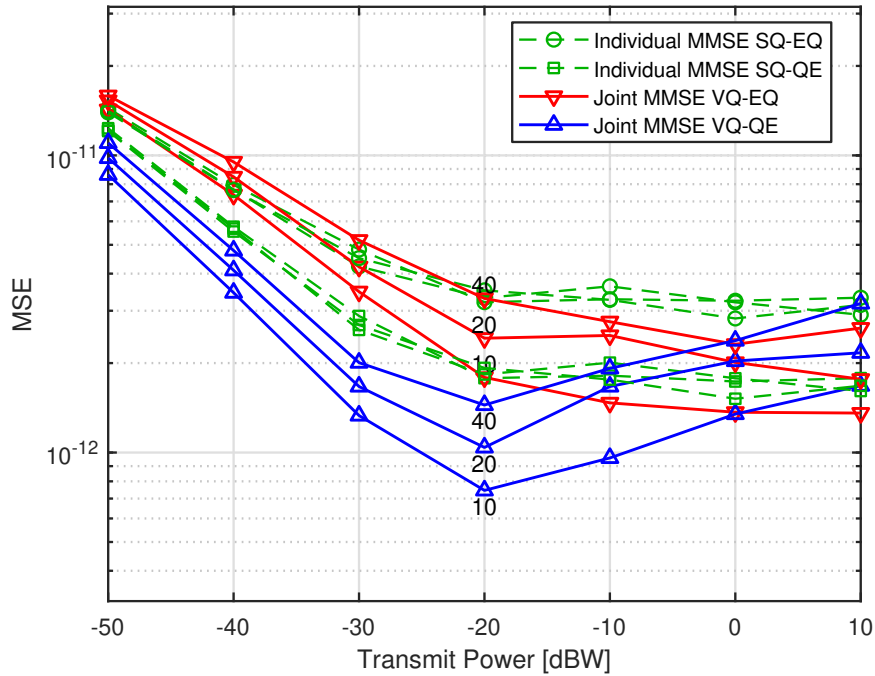}
\caption{The MSE versus transmit power for $M=120$, $N=4$, $K=20$, $C/N=2$ bits/dim and $\sigma_{\delta}=10^{\circ}, 20^{\circ}, 40^{\circ}$.}
\label{MSEvsTxPow}
\end{figure}

Fig. \ref{MSEvsTxPow} shows the MSE of different acquisition schemes against transmit power for $L=30$, $\tau=K=20$, $N=4$ and $C/N=2$ bits/dim or equivalent to the fronthaul capacity $C$ of $8$ bits. Along with VQ-EQ and VQ-QE we also present two other schemes as baselines in which uniform Scalar Quantization (SQ) and estimation are performed at the individual antennas of the APs for both EQ and QE. For each scheme we plot three curves with different angular spread standard deviation $\sigma_{\delta} = 10^{\circ}, 20^{\circ}, 40^{\circ}$, with Gaussian distributed $\delta$. It is expected that the correlation becomes weaker as $\sigma_{\delta}$ increases. The angles of arrival $\theta$ are assumed to be random uniformly distributed in $[-\pi, \pi]$ according to the distribution of users.

As can be observed in Fig. \ref{MSEvsTxPow} the VQ-EQ and VQ-QE generally can provide improvements to the baseline schemes. For both schemes the channel estimate becomes more accurate as the channel correlation increases, which is not the case for the baseline schemes. As the transmit power increases up to $-20 \text{ dB}$ all schemes show improving performance as expected. It should be noted that in our simulation set-up the path losses are large which leads to small channel gains and small typical values of MSE. In all cases, the lowest MSE can be achieved by VQ-QE at $-20 \text{ dB}$ transmit power when strong spatial correlation is present. Above this power the MSE performance of the other schemes remains constant, but that of VQ-QE degrades. In this regime the quantization noise more dominates the additve noise so that increasing the transmit power has little effect.

\begin{figure}[ht!]\centering
\includegraphics[width=0.7\columnwidth]{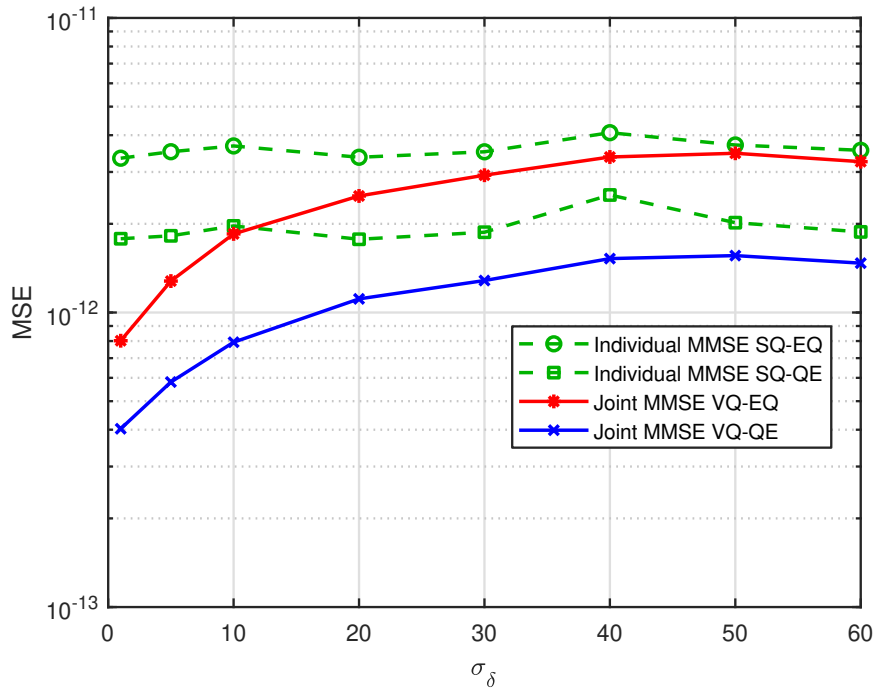}
\caption{The MSE versus angular spread standard deviation $\sigma_{\delta}$ for Gaussian distributed $\delta$, $M=120$, $K=20$, $C/N=2$ bits/dim, and TxPower=-20dB}
\label{MSEvsTheta}
\end{figure}

The estimator of VQ-QE is derived based on simplifying assumption that the quantizer input and the respected quantization noise is Gaussian. Below $-20 \text{ dB}$ the additive noise is still able to make the effective noise like Gaussian so that the estimator for VQ-QE performs quite well. But in the regime above $-20\text{ dB}$ the correlated quantization noise accros the antennas is considerably not Gaussian which leads to mismatch in the estimation. Similar behaviour is also observed in \cite{Kim2018ChannelEF}, where one-bit channel estimation is performed for co-located massive MIMO with spatio and temporal correlation. The effect is also explained by the mismatch of the quantization noise to the Gaussian assumption of the estimator: this is more significant above -20 dB.  A full discussion of this effect is however beyond the scope of the present paper. Although the MSE of VQ-QE increases at higher transmit power, it is still roughly equal to the SQ-EQ scheme in the asymptotic regime.

Fig. \ref{MSEvsTheta} shows the dependence of MSE on spatial correlation (i.e. $\sigma_{\delta}$). This figure confirms that the proposed schemes, at least at moderate SNR, can effectively exploit the strong channel correlation to achieve more accurate CSI. In the asymptotic regime, where the channels are uncorrelated, the VQ schemes can still achieve a considerable gain due to the space filling advantage obtained from the dense packing codebooks of VQ.

\begin{figure}[ht!]\centering
\includegraphics[width=0.7\columnwidth]{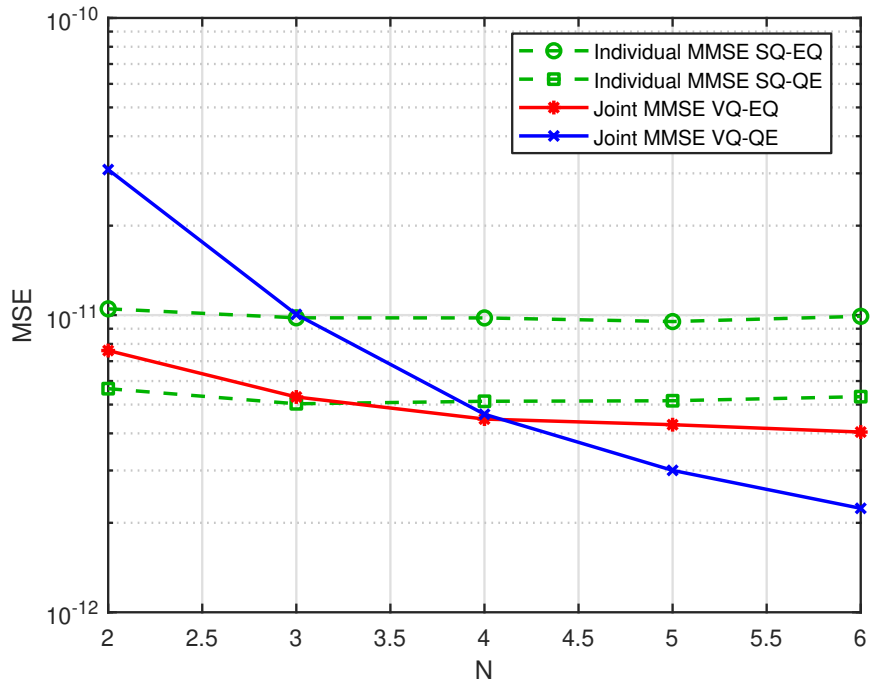}
\caption{The MSE versus number of antennas per AP $N$ for $M=120$, $K=20$, $C/N=1$ bits/dim, TxPower=-20dB and $\sigma_{\delta}=10^{\circ}$.}
\label{MSEvsN}
\end{figure}

For further evaluation, we investigate in Fig. \ref{MSEvsN} the relationship of the MSE to the number of antennas per AP. We let the number of antennas $N$ increase whereas the number of bits per antenna and the total number of antennas are fixed respectively to $C/N=1$ and $M=120$. In this case, increasing $N$ means also increasing the fronthaul capacity $C$. As we can observe in Fig. \ref{MSEvsN} that the MSE performance of the baseline schemes is independent of N. In contrast, vector quantization, especially VQ-QE, is able to exploit the correlation of the antennas at an AP to improve the CSI accuracy.

\section{Conclusion} \label{Conclusion}
We have presented in this paper the advantage of utilizing vector quantization for CSI acquisition at the CPU of cell-free massive MIMO under limited fronthaul capacity with spatial channel correlation. We investigated the performance of few-bit vector quantization for two different CSI acquisition strategies, VQ-EQ and VQ-QE. For the first time in this paper we derived an estimator based on the Bussgang theorem for vector quantization and evaluated the performance numerically. Our results show that the VQ-QE gives the best performance in terms of MSE under the condition of strong spatial correlation and moderate SNR regime compared to other schemes.


\footnotesize{
\bibliographystyle{IEEEtran}
\bibliography{References}
\nocite{*}
}
\IEEEpeerreviewmaketitle

\end{document}

%% file: modelIlustration.tex
\usetikzlibrary{shapes}

\definecolor{pinegreen}{cmyk}{0.92,0,0.59,0.25}
\definecolor{royalblue}{cmyk}{1,0.50,0,0}
\definecolor{lavander}{cmyk}{0,0.48,0,0}
\definecolor{violet}{cmyk}{0.79,0.88,0,0}
\tikzstyle{cblue}=[circle, draw, thin,fill=cyan!20, scale=0.8]
\tikzstyle{qgre}=[rectangle, draw, thin,fill=green!20, scale=0.8]
\tikzstyle{rpath}=[ultra thick, red, opacity=0.4]
\tikzstyle{legend_isps}=[rectangle, rounded corners, thin, 
                       fill=gray!20, text=blue, draw]
                        
\tikzstyle{legend_overlay}=[rectangle, rounded corners, thin,
                           top color= white,bottom color=green!25,
                           minimum width=2.5cm, minimum height=0.8cm,
                           pinegreen]
\tikzstyle{legend_phytop}=[rectangle, rounded corners, thin,
                          top color= white,bottom color=cyan!25,
                          minimum width=2.5cm, minimum height=0.8cm,
                          royalblue]
\tikzstyle{legend_general}=[rectangle, rounded corners, thin,
                          top color= white,bottom color=lavander!25,
                          minimum width=2.5cm, minimum height=0.8cm,
                          violet]
                          
\tikzset{naming/.style={align=center,font=\small}}
\tikzset{antenna/.style={insert path={-- coordinate (ant#1) ++(0,0.2) -- +(135:0.15) + (0,0) -- +(45:0.15)}}}
\tikzset{station/.style={naming,draw,shape=dart,shape border rotate=90, minimum width=10mm, minimum height=10mm,outer sep=0pt,inner sep=3pt}}
\tikzset{mobile/.style={naming,draw,shape=rectangle,minimum width=12mm,minimum height=6mm, outer sep=0pt,inner sep=3pt}}
\tikzset{radiation/.style={{decorate,decoration={expanding waves,angle=90,segment length=4pt}}}}

\newcommand{\UE}[1]{%
\begin{tikzpicture}[every node/.append style={rectangle, scale=0.65, minimum height=1.5em, minimum width=0.5em, node distance=5cm, thick,}]
\node[mobile] (box) {#1};
\draw (box.north) [antenna=1];

\fill (box.south west) 
      (box.south east) ;

\end{tikzpicture}
}

\newcommand{\CPU}[1]{%
\begin{tikzpicture}[every node/.append style={cloud, fill=gray!20, cloud ignores aspect, cloud puffs=26, cloud puff arc= 20,  minimum width=3cm, minimum height=1.5cm, aspect=1}
\node[cpu] (box) {#1};

\fill (box.south west) 
      (box.south east) ;

\draw (box.north) [];

\end{tikzpicture}
}

\newcommand{\MBS}[1]{%
\begin{tikzpicture}[every node/.append style={rectangle, scale=0.8,  minimum height=0.5em, minimum width=0.5em, node distance=0.5cm, very thick,}]

\node[station, fill=blue!20] (base) {#1};


\draw[line cap=rect] ([xshift=-.1768cm,yshift=.6pt]base.north -| base.right tail) [antenna=1];
\draw[line cap=rect] ([yshift=.6pt]ant1 |- base.north) -- node[above,shape=rectangle,inner ysep=+.1em]{} ([xshift=.1768cm,yshift=.6pt]base.north -| base.left tail) [antenna=2];

\end{tikzpicture}
}
\resizebox{0.65\columnwidth}{!}{%
\begin{tikzpicture}[auto, thick]
  \node[cloud, name=cpu, fill=gray!20, cloud puffs=16, cloud puff arc= 100,
        minimum width=3cm, minimum height=1.5cm, aspect=1] at (-0.2,4.5) {CPU};

  \foreach \place/\x in {{(5.5,0.3)/1}, {(2.75,-0.75)/2},{(-3.25,-0.85)/3},
{(5.25,3.5)/4}, {(-0.5,1.5)/5},{(-4.2,4.0)/6}, {(-6,2.0)/7},
{(5,8.3)/8}, {(1.5,7.5)/9},{(-1.2,8.55)/10}, {(-5.5,7.5)/11}}
  \node[] (a\x) at \place {\MBS{AP}};

  \foreach \place/\y in {{(3.2,1.4)/1}, {(4.75,-1.5)/2},{(-1.5,-0.75)/3},
{(2.85,2.95)/4}, {(-3.25,0.75)/5},{(-2.2,2.5)/6}, {(-4.7,2.0)/7},
{(3,8.25)/8}, {(3.95,4.95)/9},{(-1.25,6.5)/10}, {(-3.5,7.5)/11}}
  \node[] (b\y) at \place {\UE{UE}};

 \foreach \place/\z in {{(3,-2.75)/1}, {(4.5,-2.75)/2},{(-1.5,-2.75)/3},{(0,-2.75)/4}}
  \node[] (c\z) at \place {};

\path[thin] (c1) edge[thick, gray, yshift=-2em] (c2);
 \path[thick, ->] (c3) edge[thick, dashed,red, yshift=-2em] (c4);
\node[] at (5.75, -2.75){ : Fronthaul link};
\node[] at (1.0, -2.75){ : Radio link};

\path[thin] (cpu) edge[thick, gray, yshift=-2em] (a1);
\path[thin] (cpu) edge[thick, gray, yshift=-2em] (a2);
\path[thin] (cpu) edge[thick, gray, yshift=-2em] (a3);
\path[thin] (a4) edge[thick, gray, yshift=-2em] (cpu);
\path[thin] (a5) edge[thick, gray, yshift=-2em] (cpu);
\path[thin] (a6) edge[thick, gray, yshift=-2em] (cpu);
\path[thin] (a7) edge[thick, gray, yshift=-2em] (cpu);
\path[thin] (a8) edge[thick, gray, yshift=-2em] (cpu);
\path[thin] (a9) edge[thick, gray, yshift=-2em] (cpu);
\path[thin] (a10) edge[thick, gray, yshift=-2em] (cpu);
\path[thin] (a11) edge[thick, gray, yshift=-2em] (cpu);  

\path[thick, ->] (b4) edge[thick, dashed,red, yshift=-2em] (a1);
\path[thick, ->] (b4) edge[thick, dashed,red, yshift=-2em] (a2);
\path[thick, ->] (b4) edge[thick, dashed,red, yshift=-2em] (a4);
\path[thick, ->] (b4) edge[thick, dashed,red, yshift=-2em] (a5);
\path[thick, ->] (b4) edge[thick, dashed,red, yshift=-2em] (a8);
\path[thick, ->] (b4) edge[thick, dashed,red, yshift=-2em] (a9);

\end{tikzpicture}
}